\documentstyle[preprint,tighten,aps,eqsecnum,epsf,12pt]{revtex}
\makeatletter

\providecommand{\LyX}{L\kern-.1667em\lower.25em\hbox{Y}\kern-.125emX\@}
\let\SF@@footnote\footnote
\def\footnote{\ifx\protect\@typeset@protect
    \expandafter\SF@@footnote
  \else
    \expandafter\SF@gobble@opt
  \fi
}
\expandafter\def\csname SF@gobble@opt \endcsname{\@ifnextchar[
  \SF@gobble@twobracket
  \@gobble
}
\edef\SF@gobble@opt{\noexpand\protect
  \expandafter\noexpand\csname SF@gobble@opt \endcsname}
\def\SF@gobble@twobracket[#1]#2{}

\newcommand{\lyxaddress}[1]{
  \par {\raggedright #1 
  \vspace{1.4em}
  \noindent\par}
}

\makeatother

\begin{document}

\preprint{\begin{minipage}{2in}\begin{flushright}
  TRI-PP-01-08 \end{flushright}
  \end{minipage}}

\title{Charmed baryons in lattice QCD}

\author{Randy Lewis$^1$, N. Mathur$^2$ and R.M. Woloshyn$^2$}

\maketitle

\lyxaddress{\centering $^1$Department of Physics, University of Regina, Regina,
SK, Canada S4S 0A2\\
$^2$TRIUMF, 4004 Wesbrook Mall, Vancouver BC, Canada V6T 2A3}

\begin{abstract}
Masses of singly and doubly charmed baryons are calculated in quenched lattice
QCD using an improved action of the D234 type on an anisotropic lattice. The
mass differences between spin 3/2 and spin 1/2 baryon states are calculated
and compared to mass differences between vector and pseudoscalar mesons. The
suppression of spin splittings in mesons containing heavy quarks, characteristic
of quenched QCD simulations, is not observed in the baryon sector. The mass
dependence of color hyperfine effects is discussed within the context of the
quark model and heavy quark effective theory.
\end{abstract}
\thispagestyle{empty}\newpage

\section{INTRODUCTION}

Quenched lattice quantum chromodynamics (QCD) does reasonably well at describing
hadronic phenomena. However there are a few well established instances where
quenched approximation clearly fails to reproduce experimental values. One well
studied example is the persistent underestimate of mass differences between
vector and pseudoscalar mesons containing heavy quarks \cite{bernard}. While
the situation is still unsettled, the first attempts with unquenched simulations
do not yet resolve this discrepancy\cite{koniuk,manke,collins} so, clearly,
a better understanding of color hyperfine effects in lattice QCD would be
helpful.
One way to proceed is to explore the situation in heavy baryons. Unfortunately,
this is not so easy since spin splittings in the baryon sector are smaller than
in mesons while lattice QCD correlators for baryons (especially spin 3/2 states)
are noisier than those of mesons. An early work \cite{UKQCD1} suggested that
spin splittings in heavy baryons are indeed very small, but later studies for
charmed baryons \cite{cbary} and for bottom baryons \cite{ali1} obtained results
more in line with phenomenological expectations.

In this paper we present further results for charmed baryons using the approach
of \cite{cbary}. Calculations were carried out for lattices with different
values of the lattice spacing. As well, the physical size of the lattice was
increased compared to that used in \cite{cbary}.

Section II presents some details of the lattice simulation. The calculations
are done with an improved action of the D234 type \cite{alford} on an
anisotropic
lattice. This action has been used and described previously so the detailed
expressions are relegated to an Appendix. The results are given in Section III
along with some discussion about the fitting procedure and about the estimate
of systematic errors. No attempt was made in the present work to do a continuum
extrapolation. At each lattice spacing, a number of systematic effects have been
identified and estimates were made for uncertainties induced in masses and mass
differences. These include the overall scale uncertainty, the choice of time
window for fitting correlation functions, ambiguity in fixing the strange and
charm quark mass and in the extrapolation to physical up and down quark masses.

In section IV we try to compare spin splittings in baryons with those in mesons.
It turns out that, for a subset of mesonic and baryonic states, there is a very
simple mass dependence of color hyperfine effects. In fact over the whole mass
range where experimental masses are available the ratio of meson to baryon mass
difference is remarkably constant. This fact is used as a benchmark against
which to view our lattice QCD results. It is suggested that, whereas quenched
lattice QCD underestimates mesonic spin splittings, in baryons the situation
is qualitatively different with no suppression of spin splittings being
observed.
Rather there is a tendency for spin splittings to be overestimated.

\section{METHOD}

The calculation is done on an anisotropic lattice using gauge field and quark
actions that are improved \cite{alford} by removing tree level errors up to
\( O(a^{2}) \) and introducing renormalization factors for the gauge links
to remove the dominant errors due to tadpole fluctuations. The tadpole factors
are estimated using the expectation value of the gauge field link in Landau
gauge. These actions have been used previously for heavy-light mesons 
\cite{spwave}
and in a preliminary study of charmed baryons \cite{cbary}. The expressions
for the actions are given in the appendix.

Hadron masses are calculated from zero-momentum correlation functions in the
usual way. This choice of interpolating operators for the hadrons is not unique.
For \( \Sigma  \)-like baryons containing quarks with two different flavors
(denoted by Q and q) a common choice \cite{leinweb1} is
\begin{equation}
\label{proton}
\epsilon ^{abc}[q^{T}_{a}C\gamma _{5}Q_{b}]q_{c},
\end{equation}
 where a,b,c are color indices and Dirac indices have been suppressed. Taking
\( q=u \) and \( Q=d \), this operator will give the usual interpolating
operator
for the proton and, as discussed in Ref. \cite{cbary}, it is advantageous to
use this operator for all flavor combinations as this gives a unified analysis
over the whole mass range. For the \( \Xi _{Q}' \) which has light quarks of
different flavor coupled to spin 1, the operator 
\begin{equation}
\frac{1}{\sqrt{2}}\left\{ \epsilon ^{abc}[q_{a}'^{T}C\gamma _{5}Q_{b}]q_{c}
+\epsilon ^{abc}[q^{T}_{a}C\gamma _{5}Q_{b}]q_{c}'\right\} 
\end{equation}
symmetrized in the light quarks is used. The \( \Lambda  \)-like baryons
containing
a heavy quark can be interpolated using an operator of the form 
\begin{equation}
\label{heavy}
\epsilon ^{abc}[q^{T}_{a}C\gamma _{5}q_{b}']Q_{c}.
\end{equation}
A more symmetrical choice would be the octet lambda
\begin{equation}
\label{octet}
\frac{1}{\sqrt{6}}\epsilon ^{abc}\left\{ 2[q^{T}_{a}C\gamma _{5}q_{b}']Q_{c}
+[q^{T}_{a}C\gamma _{5}Q_{b}]q_{c}'-[q_{a}'^{T}C\gamma _{5}Q_{b}]q_{c}\right\} 
\end{equation}
which is degenerate with the \( \Sigma  \) in the SU(3) flavor limit. This
choice is consistent with the idea of using operators that allow for a unified
calculation over all masses. In fact, it was found that the operators
(\ref{heavy})
and (\ref{octet}) give masses which are consistent within statistical errors
so results using (\ref{octet}) are reported in this paper.

For spin 3/2 states we use the simple operator (appropriate for the heavy quark
limit)

\begin{equation}
\label{delta}
\epsilon ^{abc}[q^{T}_{a}C\gamma _{\mu }q_{b}']Q_{c}
\end{equation}
for all states since there are indications \cite{valencia} that the more
symmetrical
form used in \cite{cbary} leads to essentially identical results.

The spin 3/2 field (\ref{delta}) propagates both spin 1/2 and spin 3/2 states
\cite{benmer}. At zero momentum the correlation function with spatial Lorentz
indices has the general form \cite{UKQCD1}

\begin{equation}
\label{corfun}
C_{ij}(t)=(\delta _{ij}-\frac{1}{3}\gamma _{i}\gamma _{j})C_{3/2}(t)+
\frac{1}{3}\gamma _{i}\gamma _{j}C_{1/2}(t),
\end{equation}
where the subscripts 3/2 and 1/2 denoted the spin projections. By choosing
different Lorentz components the quantity \( C_{3/2}(t) \) is extracted and
used to determine the mass of the spin 3/2 states. 

Hadron correlators were calculated using interpolating operators in local form
at both source and sink and also applying a gauge invariant smearing to the
quark propagators at the sink. The gauge invariant smearing function, Eq. (13)
of Ref. \cite{smear}, was used. Hadron masses were obtained by a simultaneous
fit to local and sink-smeared correlators taking into account correlations
between
different time slices with the inverse of the covariance matrix obtained using
singular value decomposition. Two exponential functions are used in the fit
to the local correlator. The sink-smeared correlation function is fit by a
single
exponential using a mass parameter constrained to be the same as the ground
state mass of the fit to the local correlator. The time window for the fit was
obtained by fixing the ending time to be sufficiently large so that the fits
were not sensitive to its value and by varying the starting time so that a
minimum
of \( \chi ^{2} \) (typically around 1/d.o.f.) was achieved. Fits with starting 
times \( \pm 1 \) time
step away from the time giving the minimum \( \chi ^{2} \) were used to estimate
the systematic uncertainty associated with fixing the fitting time window.

\section{Results}

Calculations were carried out for three different sets of quenched gauge
configurations
on anisotropic lattices with a bare aspect ratio \( a_{s}/a_{t}=2 \) and spatial
lattice spacing varying from about 0.22fm to 0.15fm. Gauge fields were
constructed
using pseudo-heatbath Monte Carlo with 400 \( (\beta =2.1) \) to 800 
\( (\beta =2.5) \)
sweeps between saved configurations. Fixed time boundaries were used in
calculating
the quark propagators. The parameters for the lattices are given in Table 
\ref{latparms}. 

Six values of the hopping parameter were used and are listed in Table 
\ref{hopparms}.
The smallest values were chosen in the region of the charm quark and the four
larger values were used for extrapolating to the light (u,d) quark mass. The
temporal
lattice spacing was fixed by calculating the \( \rho  \)-meson mass. As well
known, quenched lattice QCD does not give a mass spectrum in perfect agreement
with experiment so alternate ways to fix the scale would lead to different
values.
This is an intrinsic limitation of the quenched approximation. The hopping
parameters
corresponding to the strange and charm quark masses were fixed from the 
\( \phi  \)-meson
and D-meson respectively. These are the values given in Table \ref{hopparms}.
A systematic uncertainty in these quark mass determinations was estimated by
also using the kaon and \( J/\psi  \) masses.

The masses of hadrons containing up and down quarks have to be extrapolated
into the light quark mass region. This is done by extrapolating the masses
calculated
at the four largest hopping parameters as a function of pion mass using the
function \( c_{0}+c_{2}m_{\pi }^{2}+c_{3}m_{\pi }^{3} \). In some cases the
mass to be extrapolated appears to be described very well by a function without
an \( m_{\pi }^{3} \) term. In such cases, where the coefficient \( c_{3} \)
comes out to be not statistically significant, the difference between aquadratic
extrapolation and a quadratic plus cubic extrapolation is included as a
contribution to the systematic error.

Some representative results of the mass determination are shown in terms of
so-called effective mass plots for singly heavy baryons in Figs. 1 and 2. The
effective mass \( M(t) \) is \( ln(g(t)/g(t+1)) \) where \( g(t) \) is the
zero-momentum time correlation function for the hadron. The solid line in the
figures shows the ground state mass obtained by a simultaneous fit to the local
and smeared correlation functions plotted over the time window of the fit, and
the dashed lines indicate the bootstrap error. Note that a two exponential form
was used to fit the local correlators. Corresponding results for doubly heavy
baryons are given in Figs. 3 and 4. 

The results for singly charmed baryons are given in Table \ref{cbmasses}. The
first error is the statistical error which is calculated using a bootstrap
analysis
employing a bootstrap sample size equal to the configuration sample size. The
second error is the combined systematic error. This includes the overall
scale uncertainty, the uncertainty due to choice of the correlation function
fitting range as well as strange and charm quark mass uncertainties and light
(u,d) quark mass extrapolation ambiguity, where applicable. For comparison,
the experimental values are also shown, where they are known. 

The masses of doubly charmed baryons were also calculated and are given in Table
\ref{ccbmasses}. No doubly heavy baryons have been observed yet in experiments
but one may hope that this situation will change in the not too distant future
(see Ref. \cite{kiselev} for a review of the possibilities for experimental
observation). Note also that the spin splittings for the doubly charmed baryons
are as large and perhaps even larger than in the singly charmed sector.

\section{DISCUSSION}

In order to see how our results for charmed baryons and particularly for the
spin splittings fit into the overall scheme of hadron physics, it is useful to
start with a slight digression and consider the spin splittings of pseudoscalar
and vector mesons. It is well known that for vector(\( V \)) and 
pseudoscalar(\( P \)) meson
pairs of the form \( Q\overline{q} \), where \( q \) is up or down and \( Q \)
is any light or heavy flavor, the squared mass difference 
\( M_{V}^{2}-M_{P}^{2} \)
is approximately constant for all flavors \( Q. \) This mass relation was first
derived within the framework of string-like models for hadrons imposing the
constraints of chiral symmetry\cite{stringy,lewellen,beane}. It can also be
understood in the quark model with a linear confining potential \cite{frank}
and, for heavy-light mesons, from heavy quark effective theory(HQET)\cite{wise}.
For our purposes it is useful to factorize the squared mass difference and make
a plot of the spin splitting \( \Delta M_{mes}=M_{V}-M_{P} \) versus the inverse
of the average meson mass \( M_{ave}=(M_{V}+M_{P})/2 \). This is given in Fig.
5 where the line is the best linear fit to the experimental points shown by
triangles. The mesons pairs included in the plot are \( (\rho ,\pi ), \) 
\( (K^{*},K), \)
\( (D^{*},D) \) and \( (B^{*},B). \) The remaining points in Fig. 5 are the
results from quenched lattice QCD calculations. The squares are results of the
present simulation and the circles are representative results collected from
a variety of published papers \cite{ali1,spwave,mescalc,mbcalc}.  The
underestimate
of color hyperfine effects by quenched lattice QCD can be seen clearly.

We would like to have a similar global view of spin splittings in the baryon
sector. In fact the string models that predict the meson squared mass difference
relation give a similar relation for baryons\cite{lewellen}. However, if one
plots the baryon mass difference \( \Delta M_{bar}=M_{3/2}-M_{1/2} \) versus
the inverse of the average baryon mass \( (M_{3/2}+M_{1/2})/2 \) for the well
measured baryon pairs \( (\Delta ,N) \), \( (\Sigma ^{*},\Sigma ) \) and 
\( (\Sigma _{c}^{*},\Sigma _{c}) \),
it is found empirically that the relation is not linear. On the other hand,
if one uses the average meson mass (associating mesons and baryons with the
same flavor content) an almost exact linear relation is found. This is shown
in Fig. 6 where the triangles are experimental data and the line is the best
linear fit. The results of the present simulation as well as some results of
published quenched lattice calculations \cite{mbcalc,barcalc} are also shown
in Fig. 6. Unfortunately, masses in the baryon sector are not determined with
the same precision as for mesons but a clear qualitative difference from Fig.
5 can be seen. No suppression of the spin splitting is evident. Although not
completely conclusive, there may be a tendency for baryon spin splittings to
be overestimated.

The linearity of the experimental results for mass differences plotted versus
a common variable in Figs. 5 and 6 implies that the ratio of mass differences
should be constant. This is shown in Fig. 7 for the experimental data where
the ratios of the meson to baryon mass differences 
\( (\rho ,\pi )/(\Delta ,N) \),
\( (K^{*},K)/(\Sigma ^{*},\Sigma ) \) and \( (D^{*},D)/(\Sigma _{c}^{*},
\Sigma _{c}) \)
are plotted versus the averaged meson mass. The ratio is constant to within
about \( 2\% \) and the solid line is the average value 2.13. This remarkable
result was anticipated by Lipkin \cite{lipkin86} (see also Lipkin and O'Donnell
\cite{lipkin97}) within the framework of a quark model although it required
a number of assumptions. For mesons and baryons with a single heavy quark, HQET
also implies a constancy in the meson to baryon spin splitting but the effective
theory can not predict the value of the ratio.

One would like to see what lattice QCD predicts for the meson to baryon mass
difference
ratio. Unfortunately the errors associated with extrapolation preclude a very
precise determination. For this reason we choose to present unextrapolated
results,
fixing the ``light'' quark at a kappa value near that of the strange quark.
The results for \( Qqq-baryons \) (q-mass fixed, Q-mass variable) are shown
in Fig. 8. The calculated ratio is quite constant especially for the simulations
at higher \( \beta  \) where the results are more precise. The fact that the
ratio falls below the empirical value, shown by the solid line in Fig. 8, is
an indication that the suppression of spin splittings is not present in baryons
in the same way as it is in mesons.

For doubly heavy baryons one can make definite predictions for the relation
between mesonic and baryonic spin splittings. Consider the simplest possible
quark model in which only quark masses and the color hyperfine interaction term
are taken into account \cite{halzen,dgh}. The operators for meson and baryon 
mass are

\begin{equation}
M_{mes}=m_{1}+m_{2}+a(\overrightarrow{\sigma _{1}}\cdot 
\overrightarrow{\sigma_{2}})/m_{1}m_{2},
\end{equation}
 and

\begin{equation}
\label{mbar}
M_{bar}=m_{1}+m_{2}+m_{3}+\frac{1}{2}\sum _{i>j}a_{ij}'(\overrightarrow{
\sigma _{i}}\cdot \overrightarrow{\sigma _{j}})/m_{i}m_{j.}
\end{equation}
 Note the factor of \( 1/2 \) in the last term of (\ref{mbar}). This reflects
the reduction in the strength of the gluon exchange between quarks in a color
anti-triplet state relative to that between a quark and antiquark in a color
singlet. Evaluating the above expressions for mesons \( Q\overline{q} \) and
\( \Xi  \) baryons \( QQq \) one finds 

\begin{eqnarray}
M(P) & = & m_{q}+m_{Q}-\frac{3c}{m_{q}m_{Q}},\\
M(V) & = & m_{q}+m_{Q}+\frac{c}{m_{q}m_{Q}},
\end{eqnarray}

\begin{eqnarray}
M(\Xi _{QQ}) & = & m_{q}+2m_{Q}+2\left\{ \frac{c_{QQ}'}{4m_{Q}^{2}}
-\frac{c_{qQ}'}{m_{q}m_{Q}}\right\} ,\\
M(\Xi _{QQ}^{*}) & = & m_{q}+2m_{Q}+2\left\{
\frac{c_{QQ}'}{4m_{Q}^{2}}+\frac{c_{qQ}'}{2m_{q}m_{Q}}\right\}, 
\end{eqnarray}
 where the coefficients \( c \) and \( c' \) depend on the probability of
finding the interacting quarks at zero separation. The relation between the
spin splittings is

\begin{equation}
\label{quarkmod}
\Delta M_{bar}=\frac{3}{4}\frac{c_{qQ}'}{c}\Delta M_{mes}.
\end{equation}
 Now, if it is assumed that the two heavy quarks act as a single heavy compact
antitriplet color source, it is reasonable to expect that \( c_{qQ}'\approx c \)
so that

\begin{equation}
\label{qmratio}
\Delta M_{bar}\approx \frac{3}{4}\Delta M_{mes}.
\end{equation}
 This diquark picture for doubly heavy baryons can also be analyzed
in a heavy quark effective theory. This has been done in an elegant formulation 
utilizing a superflavor symmetry to relate hadrons containing a heavy vector 
diquark to those with a heavy spin-1/2 antiquark \cite{savage90}. 

Although no evidence yet exists that for doubly heavy baryons the ratio of meson
to baryon spin splittings is a constant function of quark mass our lattice
calculations
suggest that it might be so. Fig. 9 shows the ratio of unextrapolated results
for \( QQq-baryons \) (q-mass fixed, Q-mass variable). The long dashed line
in the figure is the value of the ratio obtained experimentally for the strange
quark, that is the \( K^{*}-K \) mass difference divided by the 
\( \Xi ^{*}-\Xi  \)
mass difference. The short dashed line is the prediction Eq. (\ref{qmratio}). 

Eq. (\ref{quarkmod}) has been derived explicitly for the doubly heavy 
\( \Xi _{QQ},\: \Xi _{QQ}^{*} \)
system but the same expression holds for the singly heavy baryons. In this case
there is no simple relation between \( c_{qQ}' \) and \( c \) but, since the
gluon interaction between quarks is weaker than between a quark and antiquark,
one might expect that \( c_{qQ}'<c \). This implies that spin splittings in
singly heavy baryons are smaller than in doubly heavy baryons containing the
same quark flavors. This expectation is borne out empirically for the strange
\( \Sigma 's \) and \( \Xi 's \) but given the statistical and systematic
errors there is no clear evidence from our lattice simulations for charmed
baryons.

Finally, we note that if it is assumed that the ratio \( \Delta M_{mes}/
\Delta M_{bar} \)
is constant as a function of quark mass for doubly heavy baryons just as it
is for singly heavy baryons, then using the experimental values of the 
\( \Xi  \)
and \( \Xi ^{*} \) masses along with the known meson masses yields the
phenomenological
predictions 
\[
\Xi _{cc}^{*}-\Xi _{cc}=76.6MeV,\]
 and

\[
\Xi _{bb}^{*}-\Xi _{bb}=24.5MeV.\]

\section{SUMMARY}

Masses of charmed baryons were calculated in quenched lattice QCD using improved
gluon and quark actions on an anisotropic lattice. The actions were improved
to remove tree level errors up to \( O(a^{2}) \) and tadpole factors, estimated
by using gauge field links in Landau gauge, were introduced to remove the
dominant errors due to tadpole fluctuations.

Calculations were done at three different values of the gauge coupling constant
with the spatial lattice spacing varying from 0.22fm to 0.15fm. The results
at the two largest values of the gauge coupling are compatible with each other
and are in fair agreement with experimental data where it is available. Masses
and mass differences at the smallest value of the gauge coupling (largest
lattice
spacing) tend to be smaller, perhaps reflecting larger finite lattice spacing
errors.

The main focus of our work was the mass differences due to spin dependent
interactions.
It is well established that in quenched lattice QCD the mass differences between
vector and pseudoscalar mesons are underestimated for mesons containing heavy
quarks. The present simulations show no comparable suppression of splittings
between spin 1/2 and spin 3/2 baryons. Our results and results taken from the
literature indicate a tendency for baryon spin splittings to be overestimated
but, this is not established definitively due to relatively large errors
associated with baryon mass determinations.

To get an overall view of the spin splittings it was useful to consider the
scaling of mass differences with the average meson mass as one changes quark
flavors. For mesons this relation \cite{stringy} predates QCD. We have found
it useful to extend the same scaling (with average meson mass) to mass
differences
between baryons. A relation between meson and baryon spin splittings, which
is implied by this scaling, was anticipated by Lipkin \cite{lipkin86} from
a quark model analysis. From the point of view of QCD, a mass independent ratio
of meson to baryon spin splittings can be derived for heavy flavored mesons
and baryons using heavy quark effective theory, but how to extend this result
to light flavor hadrons is not clear. Nonetheless a constant meson to baryon
mass difference ratio is satisfied remarkably well by experimental data for
quark flavors from light to charm. Obviously, a definitive experimental
determination
of the \( \Sigma _{b}^{*} \) mass to extend this analysis to the b-quark region
would be extremely useful.

Our lattice QCD results are compatible with a mass independent meson to baryon
spin splitting ratio for both singly and doubly heavy baryons. However, the
more precise values obtained at our larger values of the gauge coupling indicate
a value for the ratio for singly heavy baryons smaller than the empirical value.
This can be interpreted as another indication that the suppression of spin
splittings
found for mesons does not occur for baryons. For doubly heavy baryons our
preferred
values for the meson to baryon ratio lie below the value obtained by using the
masses of the doubly strange hyperons \( \Xi  \) and \( \Xi ^{*} \). Clearly,
the experimental observation of doubly heavy baryons and the systematic
investigation
of their spin splittings would be very interesting indeed.

\section*{ACKNOWLEDGMENTS}

The authors are grateful to Mark Wise for a helpful communication.
This research was supported in part by the Natural Sciences and
Engineering Research Council of Canada.  Some of the computing
was done on hardware funded by the Canada Foundation for Innovation,
with contributions from Compaq Canada, Avnet Enterprise Solutions,
and the Government of Saskatchewan.

\appendix
\section{THE LATTICE ACTION}

The lattice action has two terms: gauge action and quark action. The entire
action is classically and tadpole-improved with the tadpole factors, \( u_{s} \)
and \( u_{t} \), defined as the mean links in Landau gauge in a spatial and
temporal direction, respectively.

The gauge field action is 
\begin{eqnarray}
S_{G}(U) & = & \frac{5\beta }{3}\left[ \frac{1}{u_{s}^{4}\xi }
\sum _{\textrm{ps}}\left( 1-\frac{1}{3}{\textrm{ReTr}}U_{\textrm{ps}}\right) 
-\frac{1}{20u_{s}^{6}\xi }\sum _{\textrm{rs}}\left( 1-\frac{1}{3}
{\textrm{ReTr}}U_{\textrm{rs}}\right) \right. \nonumber \\
 &  & +\frac{\xi }{u_{s}^{2}u_{t}^{2}}\sum _{\textrm{pt}}\left( 1-\frac{1}{3}
{\textrm{ReTr}}U_{\textrm{pt}}\right) -\frac{\xi }{20u_{s}^{4}u_{t}^{2}}
\sum _{\textrm{rst}}\left( 1-\frac{1}{3}{\textrm{ReTr}}U_{\textrm{rst}}\right)
\nonumber \\
 & - & \left. -\frac{\xi }{20u_{s}^{2}u_{t}^{4}}\sum _{\textrm{rts}}\left( 
1-\frac{1}{3}{\textrm{ReTr}}U_{\textrm{rts}}\right) \right] ,
\end{eqnarray}
 where \( \xi \equiv a_{s}/a_{t} \) is the aspect ratio and \( \beta  \) is
the lattice gauge field coupling constant. The subscripts ``ps'' and ``rs''
denote spatial plaquettes and spatial planar 1\( \times  \)2 rectangles
respectively.
Plaquettes in the temporal-spatial planes are denoted by ``pt'', while
rectangles
with the long side in a spatial(temporal) direction are labeled by 
``rst''(``rts'').
The leading classical errors of this action are quartic in lattice spacing.

An action of the D234 type \cite{alford} is used for the quarks with
coefficients
set to their tadpole-improved classical values. Its leading classical errors
are cubic in lattice spacing: 
\begin{eqnarray}
S_{F}(\bar{q},q;U) & = & \frac{4\kappa }{3}\sum _{x,i}\left[ \frac{1}{u_{s}
\xi ^{2}}D_{1i}(x)-\frac{1}{8u_{s}^{2}\xi ^{2}}D_{2i}(x)\right] \nonumber \\
 &  & +\frac{4\kappa }{3}\sum _{x}\left[ \frac{1}{u_{t}}D_{1t}(x)-\frac{1}
{8u_{t}^{2}}D_{2t}(x)\right] \nonumber \\
 &  & +\frac{2\kappa }{3u_{s}^{4}\xi ^{2}}\sum _{x,i<j}\bar{\psi }(x)
\sigma _{ij}F_{ij}(x)\psi (x)\nonumber \\
 &  & +\frac{2\kappa }{3u_{s}^{2}u_{t}^{2}\xi }\sum _{x,i}\bar{\psi }(x)
\sigma _{0i}F_{0i}(x)\psi (x)\nonumber \\
 &  & -\sum _{x}\bar{\psi }(x)\psi (x),
\end{eqnarray}
 where \( \kappa  \) denotes the hopping parameter and

\begin{eqnarray}
D_{1i}(x) & = & \bar{\psi }(x)(1-\xi \gamma _{i})U_{i}(x)\psi (x+\hat{i})
\nonumber \\
 & + & \bar{\psi }(x+\hat{i})(1+\xi \gamma _{i})U_{i}^{\dagger }(x)\psi (x),\\
D_{1t}(x) & = & \bar{\psi }(x)(1-\gamma _{4})U_{4}(x)\psi (x+\hat{t})
\nonumber \\
 & + & \bar{\psi }(x+\hat{t})(1+\gamma _{4})U_{4}^{\dagger }(x)\psi (x),\\
D_{2i}(x) & = & \bar{\psi }(x)(1-\xi \gamma _{i})U_{i}(x)U_{i}(x+\hat{i})\psi 
(x+2\hat{i})\nonumber \\
 & + & \bar{\psi }(x+2\hat{i})(1+\xi \gamma _{i})U_{i}^{\dagger }(x+\hat{i})
U_{i}^{\dagger }(x)\psi (x),\\
D_{2t}(x) & = & \bar{\psi }(x)(1-\gamma _{4})U_{4}(x)U_{4}(x+\hat{t})\psi 
(x+2\hat{t})\nonumber \\
 & + & \bar{\psi }(x+2\hat{t})(1+\gamma _{4})U_{4}^{\dagger }(x+\hat{t})
U_{4}^{\dagger }(x)\psi (x),\\
gF_{\mu \nu }(x) & = & \frac{1}{2i}\left( \Omega _{\mu \nu }(x)-
\Omega ^{\dagger }_{\mu \nu }(x)\right) -\frac{1}{3}{\textrm{Im}}\left( 
{\textrm{Tr}}\Omega _{\mu \nu }(x)\right) ,\\
\Omega _{\mu \nu } & = & \frac{-1}{4}\left[ U_{\mu }(x)U_{\nu }(x+\hat{\mu })
U_{\mu }^{\dagger }(x+\hat{\nu })U_{\nu }^{\dagger }(x)\right. \nonumber \\
 &  & +U_{\nu }(x)U_{\mu }^{\dagger }(x-\hat{\mu }+\hat{\nu })
U_{\nu }^{\dagger }(x-\hat{\mu })U_{\mu }(x-\hat{\mu })\nonumber \\
 &  & +U_{\mu }^{\dagger }(x-\hat{\mu })U_{\nu }^{\dagger }(x-\hat{\mu }
-\hat{\nu })U_{\mu }(x-\hat{\mu }-\hat{\nu })U_{\nu }(x-\hat{\nu })\nonumber \\
 &  & \left. +U_{\nu }^{\dagger }(x-\hat{\nu })U_{\mu }(x-\hat{\nu })U_{\nu }
(x+\hat{\mu }-\hat{\nu })U_{\mu }^{\dagger }(x)\right] .
\end{eqnarray}

\newpage

\begin{table}[top]

\caption{\label{latparms}Parameters for lattices used in this work.}
{\centering \begin{tabular}{cccccc}
\( \beta  \)&
size&
configurations&
\( a_{t}^{-1}(GeV) \)&
\( u_{s} \)&
\( u_{t} \)\\
\hline 
2.1&
\( 12^{3}\times 32 \)&
720&
1.803(42)&
0.7858&
0.9472\\
2.3&
\( 14^{3}\times 38 \)&
442&
2.210(72)&
0.8040&
0.9525\\
2.5&
\( 18^{3}\times 46 \)&
325&
2.625(67)&
0.8185&
0.9564\\
\end{tabular}\par}\end{table}
\begin{table}[top]

\caption{\label{hopparms}Hopping parameter values. The quantities \protect\( 
\kappa _{s}\protect \)
and \protect\( \kappa _{c}\protect \) are the hopping parameter values
associated
with strange and charm quarks respectively.}
{\centering \begin{tabular}{cccc}
\( \beta  \)&
\( \kappa 's \)&
\( \kappa _{s}(\phi ) \)&
\( \kappa _{c}(D) \)\\
\hline 
2.1&
0.175,0.176,0.229,0.233,0.237,0.240&
0.2338&
0.1739\\
2.3&
0.183.0.189,0.229,0.233,0.237,0.240&
0.2371&
0.1875\\
2.5&
0.193,0.197,0.230,0.234,0.238,0.240&
0.2382&
0.1964\\
\end{tabular}\par}\end{table}

\begin{table}[top]
\caption{\label{cbmasses}Masses of singly charmed baryons. Masses are given in
GeV, mass differences are in MeV. The first error is the statistical error and
the second is the combined systematic error. The experimental values are taken
from \protect\cite{pdg} except for \protect\( \Xi _{c}'\protect \) which is
from \protect\cite{jess}.}


{\centering \begin{tabular}{ccccc}
&
\( \beta =2.1 \)&
\( \beta =2.3 \)&
\( \beta =2.5 \)&
Experiment\\
\hline 
\( \Lambda _{c} \)&
2.272(32)\( \left( _{23}^{15}\right)  \)&
2.295(11)\( \left( _{15}^{11}\right)  \)&
2.333(20))\( \left( _{17}^{10}\right)  \)&
2.285\\
\( \Sigma _{c} \)&
2.379(31)\( \left( _{18}^{23}\right)  \)&
2.490(14)\( \left( _{33}^{17}\right)  \)&
2.493(22)\( \left( _{29}^{21}\right)  \)&
2.455\\
\( \Sigma _{c}^{*} \)&
2.440(36)\( \left( _{31}^{18}\right)  \)&
2.572(16)\( \left( _{36}^{23}\right)  \)&
2.569(26)\( \left( _{29}^{23}\right)  \)&
2.519\\
\( \Xi _{c} \)&
2.455(17)\( \left( _{42}^{11}\right)  \)&
2.462(14)\( \left( _{30}^{\, 5}\right)  \)&
2.481(14)\( \left( _{34}^{\, 1}\right)  \)&
2.468\\
\( \Xi _{c}' \)&
2.531(17)\( \left( _{35}^{11}\right)  \)&
2.594(12)\( \left( _{25}^{\, 6}\right)  \)&
2.604(13)\( \left( _{30}^{\, 8}\right)  \)&
2.560\\
\( \Xi _{c}^{*} \)&
2.583(20)\( \left( _{40}^{16}\right)  \)&
2.675(15)\( \left( _{29}^{12}\right)  \)&
2.682(15)\( \left( _{28}^{13}\right)  \)&
2.645\\
\( \Omega _{c} \)&
2.671(11)\( \left( _{59}^{11}\right)  \)&
2.699(10)\( \left( _{41}^{\, 8}\right)  \)&
2.700(11)\( \left( _{40}^{\, 8}\right)  \)&
2.704\\
\( \Omega _{c}^{*} \)&
2.722(12)\( \left( _{58}^{16}\right)  \)&
2.772(12)\( \left( _{43}^{\, 3}\right)  \)&
2.769(12)\( \left( _{40}^{\, 3}\right)  \)&
\\
\( \Sigma _{c}^{*}- \)\( \Sigma _{c} \)&
62(33)\( \left( _{32}^{19}\right)  \)&
82(12)\( \left( _{6}^{9}\right)  \)&
76(19)\( \left( _{\, 4}^{15}\right)  \)&
64\\
\( \Xi _{c}^{*}-\Xi _{c}' \)&
52(15)\( \left( _{4}^{8}\right)  \)&
82(10)\( \left( _{5}^{8}\right)  \)&
77(9)\( \left( _{5}^{7}\right)  \)&
70\\
\( \Omega _{c}^{*}-\Omega _{c} \)&
50(17)\( \left( _{\, 6}^{11}\right)  \)&
73(8)\( \left( _{5}^{7}\right)  \)&
69(7)\( \left( _{6}^{5}\right)  \)&
\\
\end{tabular}\par}\end{table}

\begin{table}[top]

\caption{\label{ccbmasses}Masses of doubly charmed baryons. Masses are given
in GeV,
mass differences are in MeV. The first error is the statistical error and the
second is the combined systematic error.}
{\centering \begin{tabular}{cccc}
&
\( \beta =2.1 \)&
\( \beta =2.3 \)&
\( \beta =2.5 \)\\
\hline 
\( \Xi _{cc} \)&
3.608(15)\( \left( _{35}^{13}\right)  \)&
3.595(12)\( \left( _{22}^{21}\right)  \)&
3.605(12)\( \left( _{19}^{23}\right)  \)\\
\( \Xi _{cc}^{*} \)&
3.666(18)\( \left( _{34}^{18}\right)  \)&
3.678(15)\( \left( _{23}^{18}\right)  \)&
3.685(14)\( \left( _{17}^{19}\right)  \)\\
\( \Omega _{cc} \)&
3.747(9)\( \left( _{47}^{11}\right)  \)&
3.727(9)\( \left( _{40}^{16}\right)  \)&
3.733(9)\( \left( _{38}^{\, 7}\right)  \)\\
\( \Omega _{cc}^{*} \)&
3.804(13)\( \left( _{54}^{18}\right)  \)&
3.800(11)\( \left( _{36}^{10}\right)  \)&
3.801(9)\( \left( _{34}^{\, 3}\right)  \)\\
\( \Xi _{cc}^{*}-\Xi _{cc} \)&
58(14)\( \left( _{10}^{16}\right)  \)&
83(8)\( \left( _{10}^{\, 7}\right)  \)&
80(10)\( \left( _{7}^{3}\right)  \)\\
\( \Omega _{cc}^{*}-\Omega _{cc} \)&
57(8)\( \left( _{\, 9}^{10}\right)  \)&
72(5)\( \left( _{5}^{4}\right)  \)&
68(5)\( \left( _{5}^{6}\right)  \)\\
\end{tabular}\par}\end{table}

\subsection*{Figure Captions}

\noindent
Fig. 1 Effective mass \( M(t) \) versus \( t \) for spin 1/2 baryons Qqq with
(a) \( \kappa _{q}=0.233,\kappa _{Q}=0.183 \) (b) \( \kappa _{q}=0.233,
\kappa _{Q}=0.189 \)
(c) \( \kappa _{q}=0.237,\kappa _{Q}=0.183 \) (d) \( \kappa _{q}=0.237,
\kappa _{Q}=0.189 \).
The solid line indicates the ground state mass and the dashed lines the
statistical error.

\medskip{}
\noindent Fig.2 Effective mass \( M(t) \) versus \( t \) for spin 3/2 baryons
Qqq with (a) \( \kappa _{q}=0.233,\kappa _{Q}=0.183 \) (b) \( 
\kappa _{q}=0.233,\kappa _{Q}=0.189 \)
(c) \( \kappa _{q}=0.237,\kappa _{Q}=0.183 \) (d) \( \kappa _{q}=0.237,
\kappa _{Q}=0.189 \).
The solid line indicates the ground state mass and the dashed lines the
statistical error.

\medskip{}
\noindent Fig.3 Effective mass \( M(t) \) versus \( t \) for spin 1/2 baryons
QQq with (a) \( \kappa _{q}=0.233,\kappa _{Q}=0.183 \) (b) \( 
\kappa _{q}=0.233,\kappa _{Q}=0.189 \)
(c) \( \kappa _{q}=0.237,\kappa _{Q}=0.183 \) (d) \( \kappa _{q}=0.237,
\kappa _{Q}=0.189 \).
The solid line indicates the ground state mass and the dashed lines the
statistical error.

\medskip{}
\noindent Fig.4 Effective mass \( M(t) \) versus \( t \) for spin 3/2 baryons
QQq with (a) \( \kappa _{q}=0.233,\kappa _{Q}=0.183 \) (b) \( 
\kappa _{q}=0.233,\kappa _{Q}=0.189 \)
(c) \( \kappa _{q}=0.237,\kappa _{Q}=0.183 \) (d) \( \kappa _{q}=0.237,
\kappa _{Q}=0.189 \).
The solid line indicates the ground state mass and the dashed lines the
statistical error.

\medskip{}
\noindent Fig. 5 The mass difference between vector and pseudoscalar meson pairs
plotted versus the inverse of the average mass. Triangles are experimental
values,
circles are results of quenched lattice calculations taken from the literature
\cite{ali1,spwave,mescalc,mbcalc} and squares the results from the present
work. The line is the best linear fit to the experimental points.

\medskip{}
\noindent Fig. 6 The mass difference between spin 1/2 and spin 3/2 baryon pairs
plotted versus the inverse of the average mass of vector and pseudoscalar mesons
with the same flavor. Triangles are experimental values, circles are results
of quenched lattice calculations taken from the literature \cite{mbcalc,barcalc}
and squares the results from the present work. The line is the best linear fit
to the experimental points.

\medskip{}
\noindent Fig. 7 The ratio of the mass difference between vector and
pseudoscalar
meson to the mass difference between spin 1/2 and spin 3/2 baryon for hadrons
with the same flavor content plotted versus average meson mass. The triangles
are experimental data for \( (\rho ,\pi )/(\Delta ,N) \), \( (K^{*},K)
/(\Sigma ^{*},\Sigma ) \)
and \( (D^{*},D)/(\Sigma _{c}^{*},\Sigma _{c}) \). The line is the average
value.

\medskip{}
\noindent Fig. 8 Lattice simulation results for the ratio of the mass difference
between vector and pseudoscalar meson to the mass difference between singly
heavy spin 1/2 and spin 3/2 baryon at \( \beta =2.1 \) (triangles), 
\( \beta =2.3 \)
(circles) and \( \beta =2.5 \) (squares). The line is the average experimental
value from Fig. 7.

\medskip{}
\noindent Fig. 9 Lattice simulation results for the ratio of the mass difference
between vector and pseudoscalar meson to the mass difference between doubly
heavy spin 1/2 and spin 3/2 baryon at \( \beta =2.1 \) (triangles), 
\( \beta =2.3 \)
(circles) and \( \beta =2.5 \) (squares). The long dashed line is the
experimental
value for \( (K^{*},K)/(\Xi ^{*},\Xi ) \). The short dashed line is the
prediction
from Eq. (\ref{qmratio}).

\epsfxsize=460pt \epsfbox[36   54  552  743]{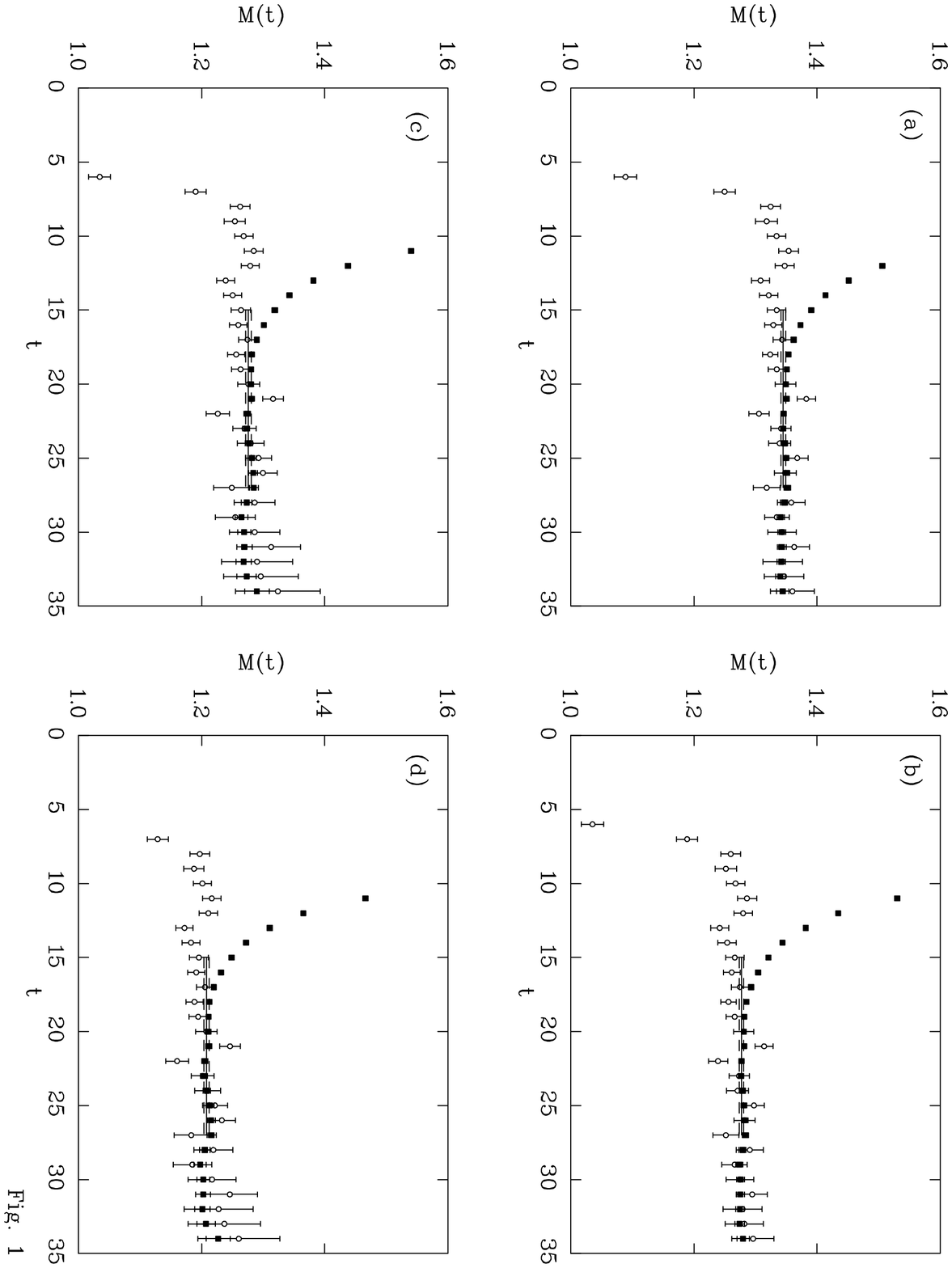}
\epsfxsize=460pt \epsfbox[36   54  552  743]{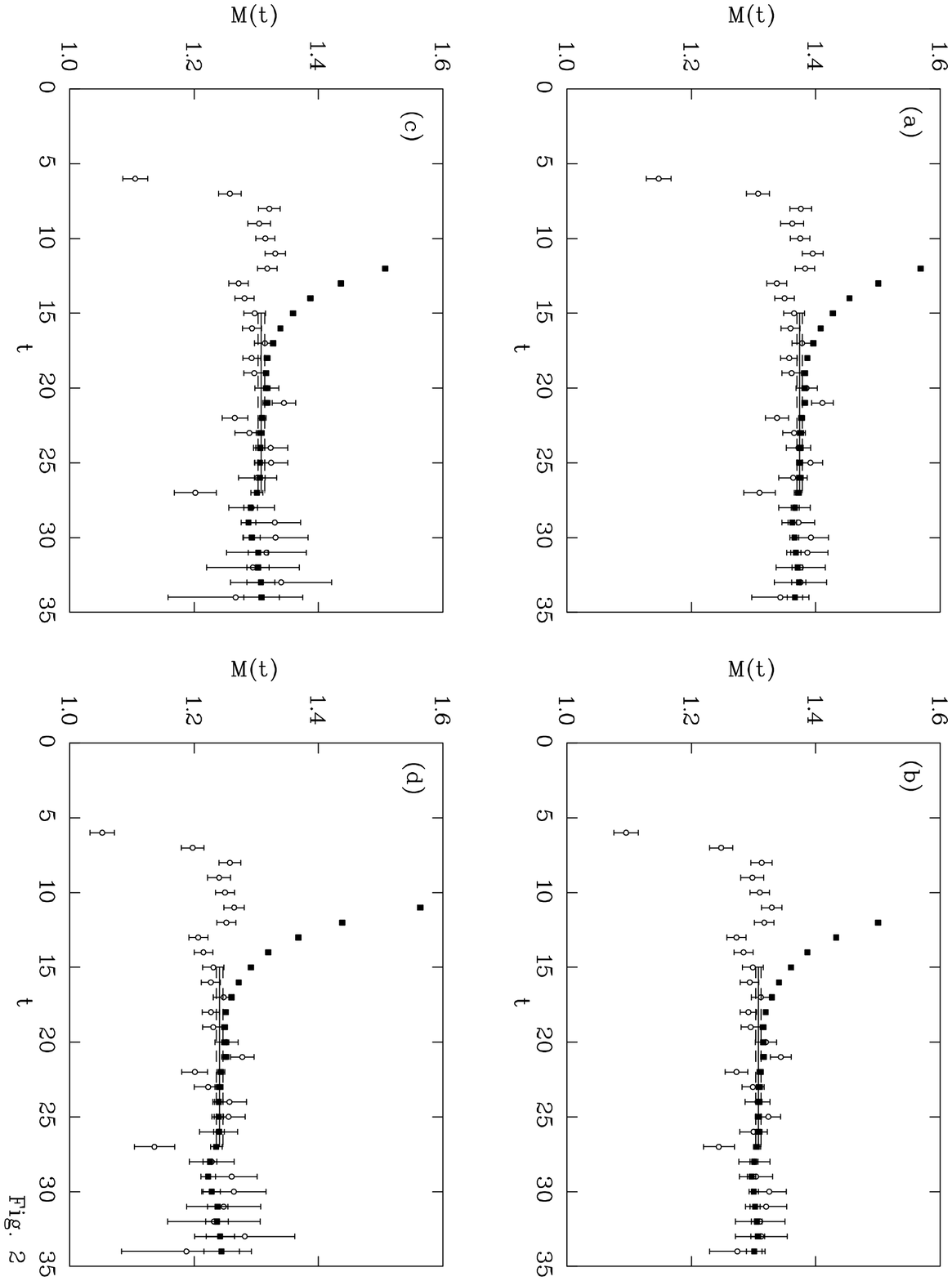}
\epsfxsize=460pt \epsfbox[36   54  552  743]{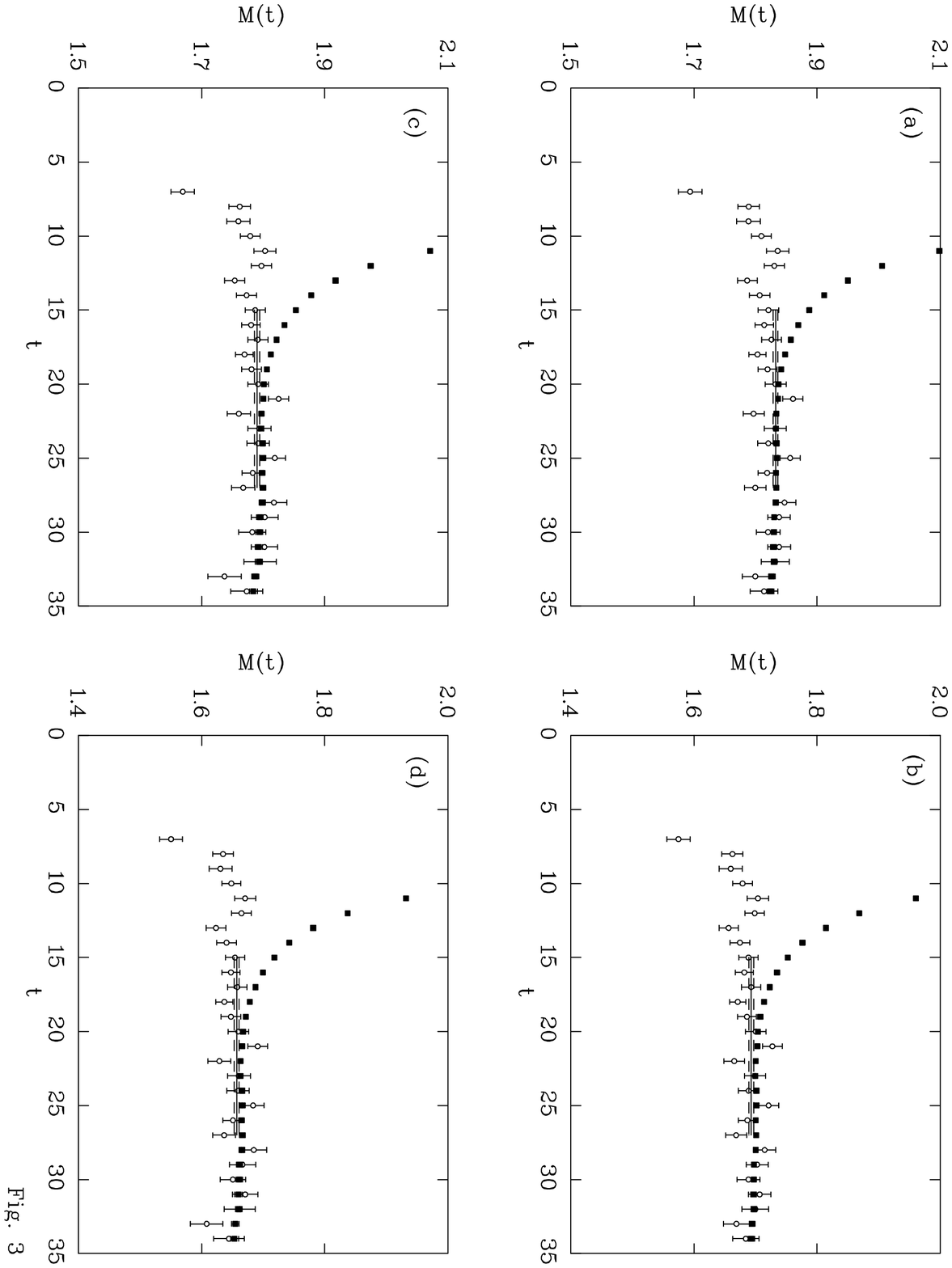}
\epsfxsize=460pt \epsfbox[36   54  552  743]{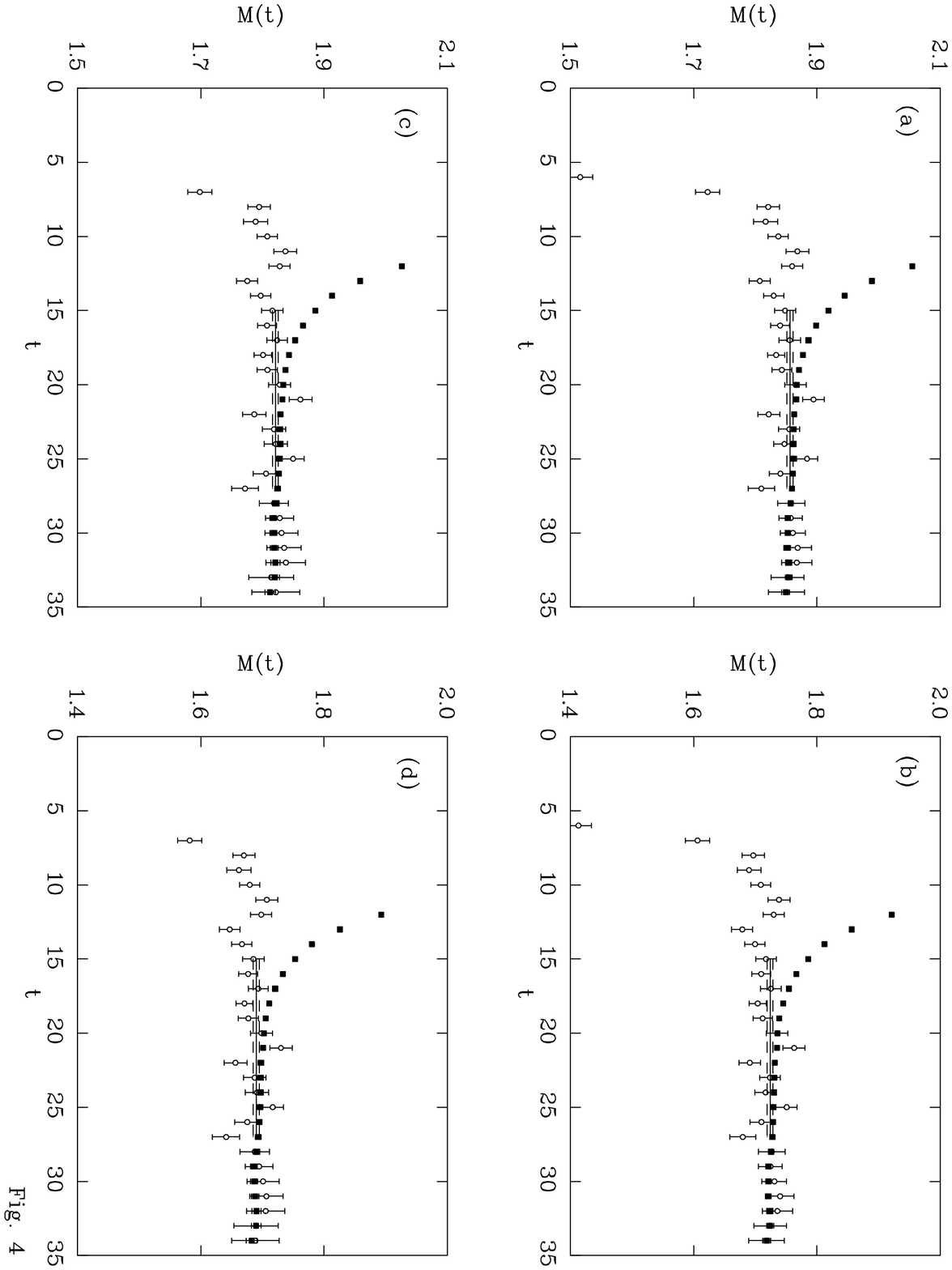}
\epsfxsize=460pt \epsfbox[36   54  552  743]{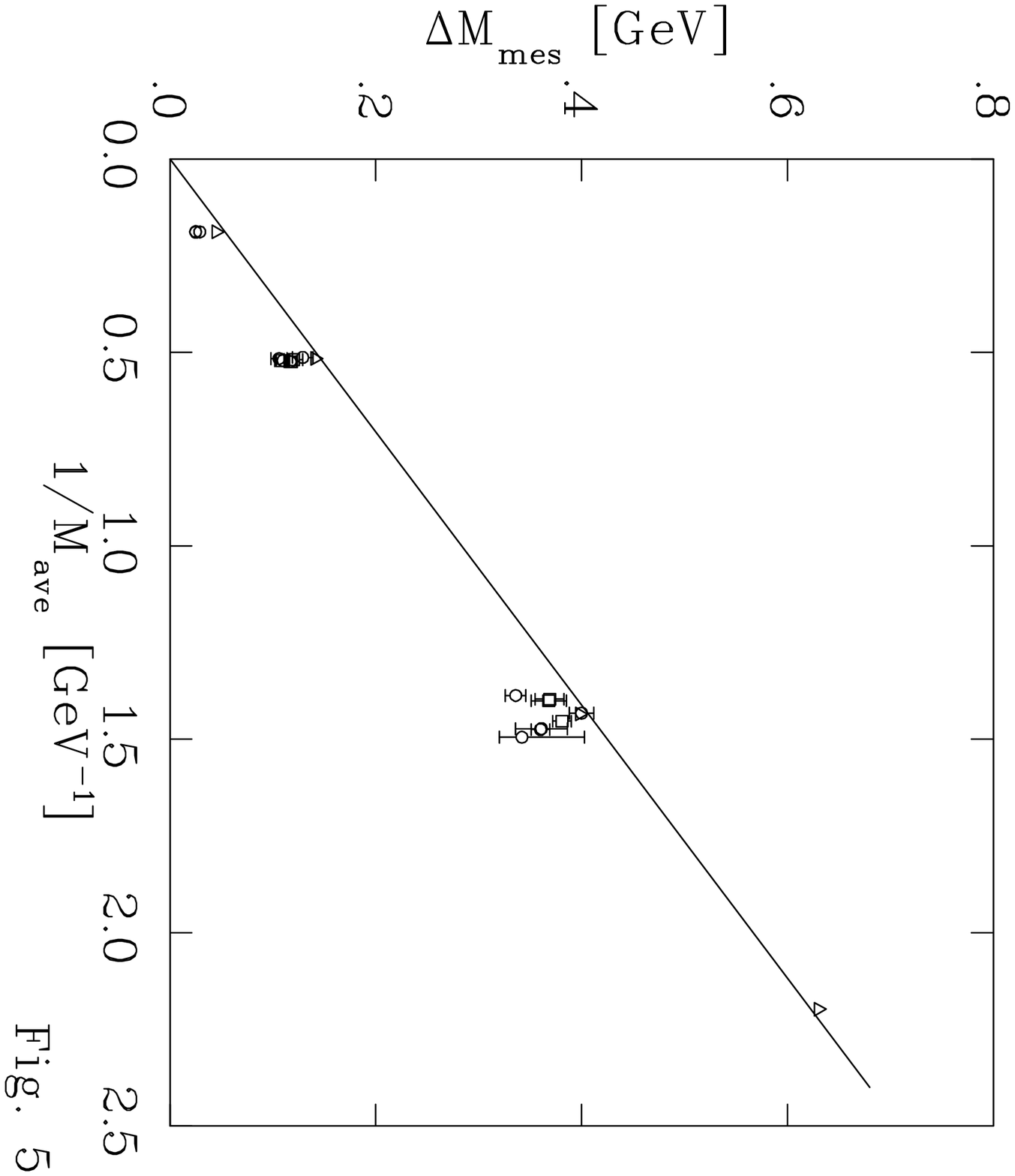}
\epsfxsize=460pt \epsfbox[36   54  552  743]{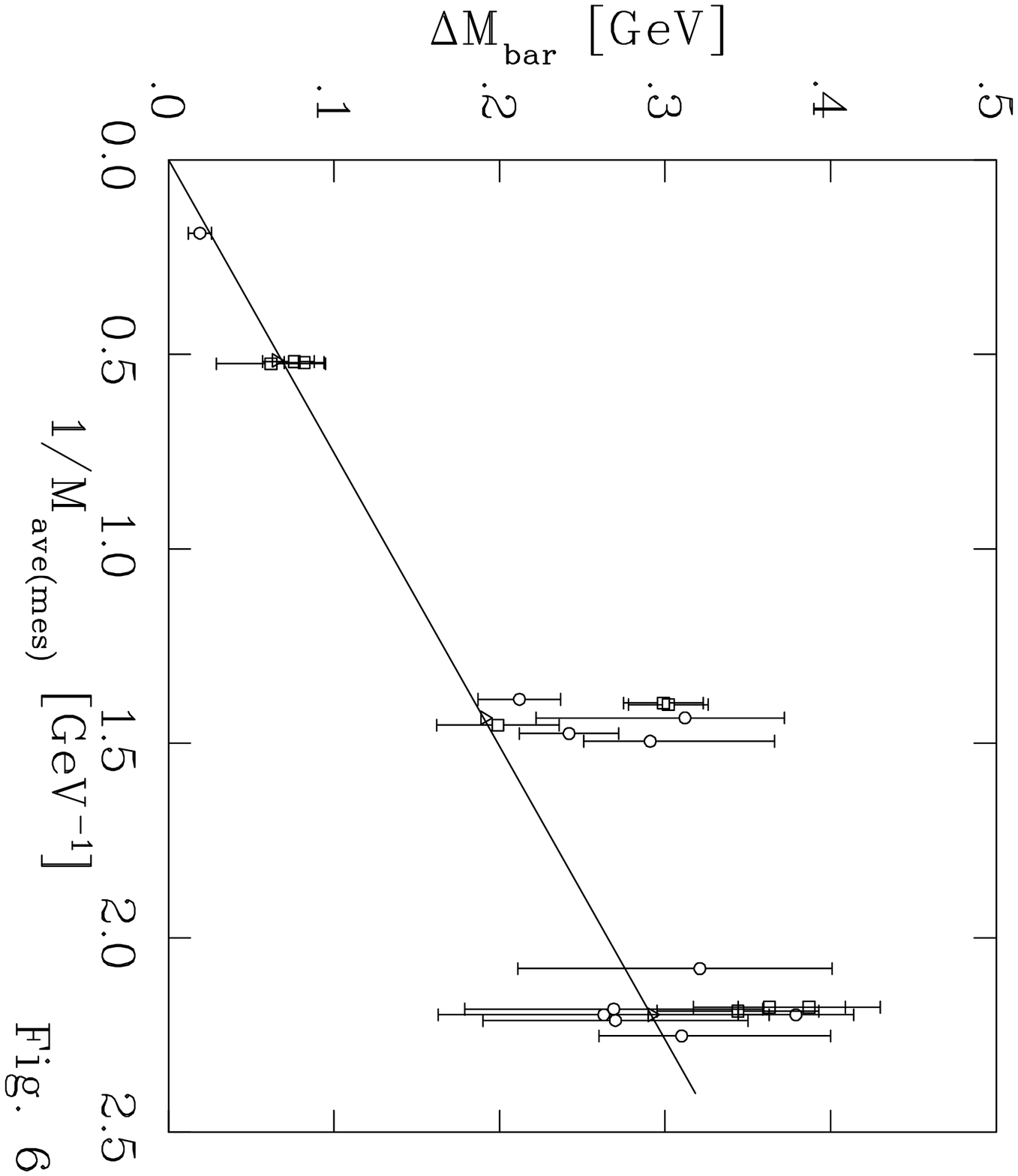}
\epsfxsize=460pt \epsfbox[36   54  552  743]{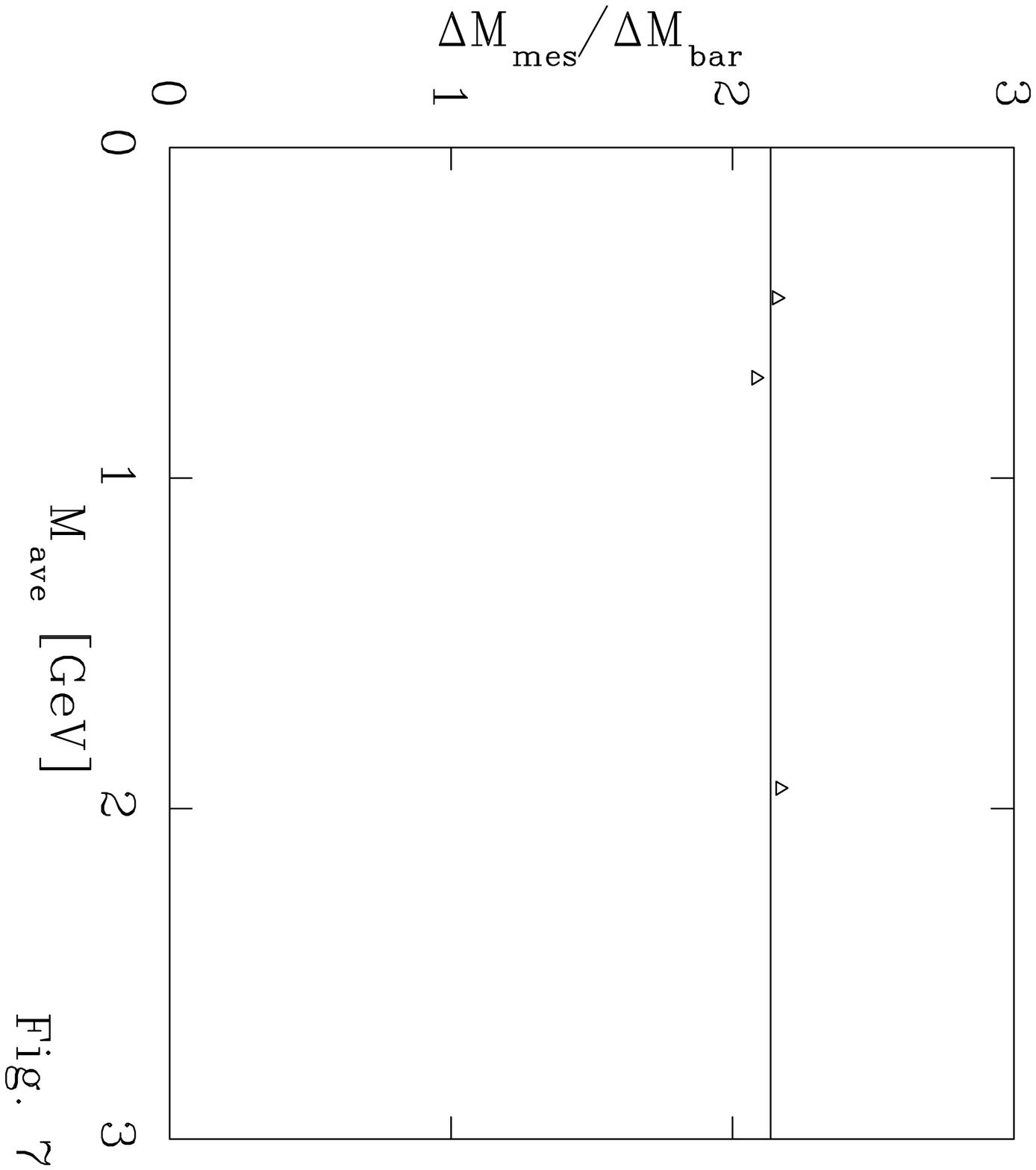}
\epsfxsize=460pt \epsfbox[36   54  552  743]{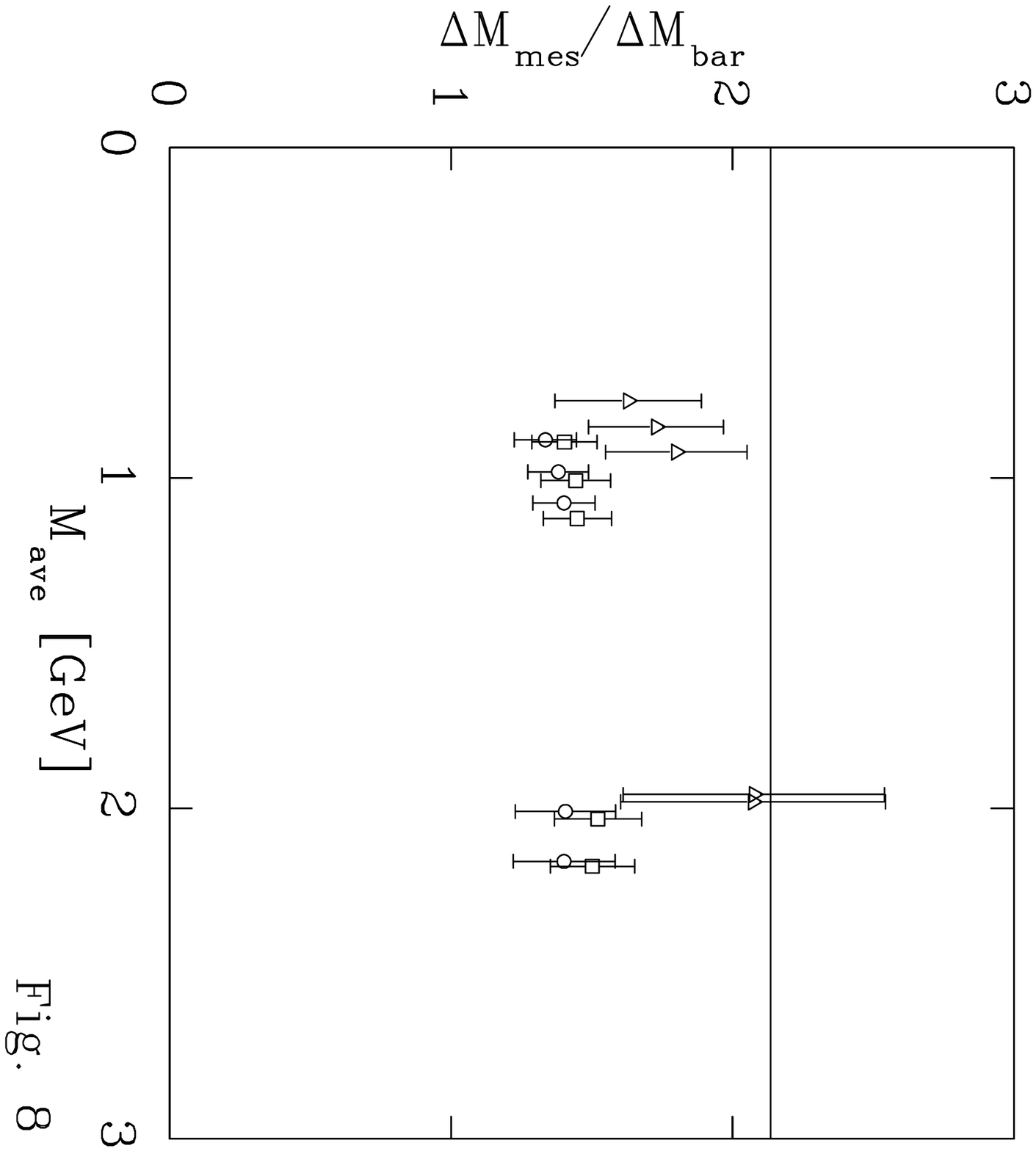}
\epsfxsize=460pt \epsfbox[36   54  552  743]{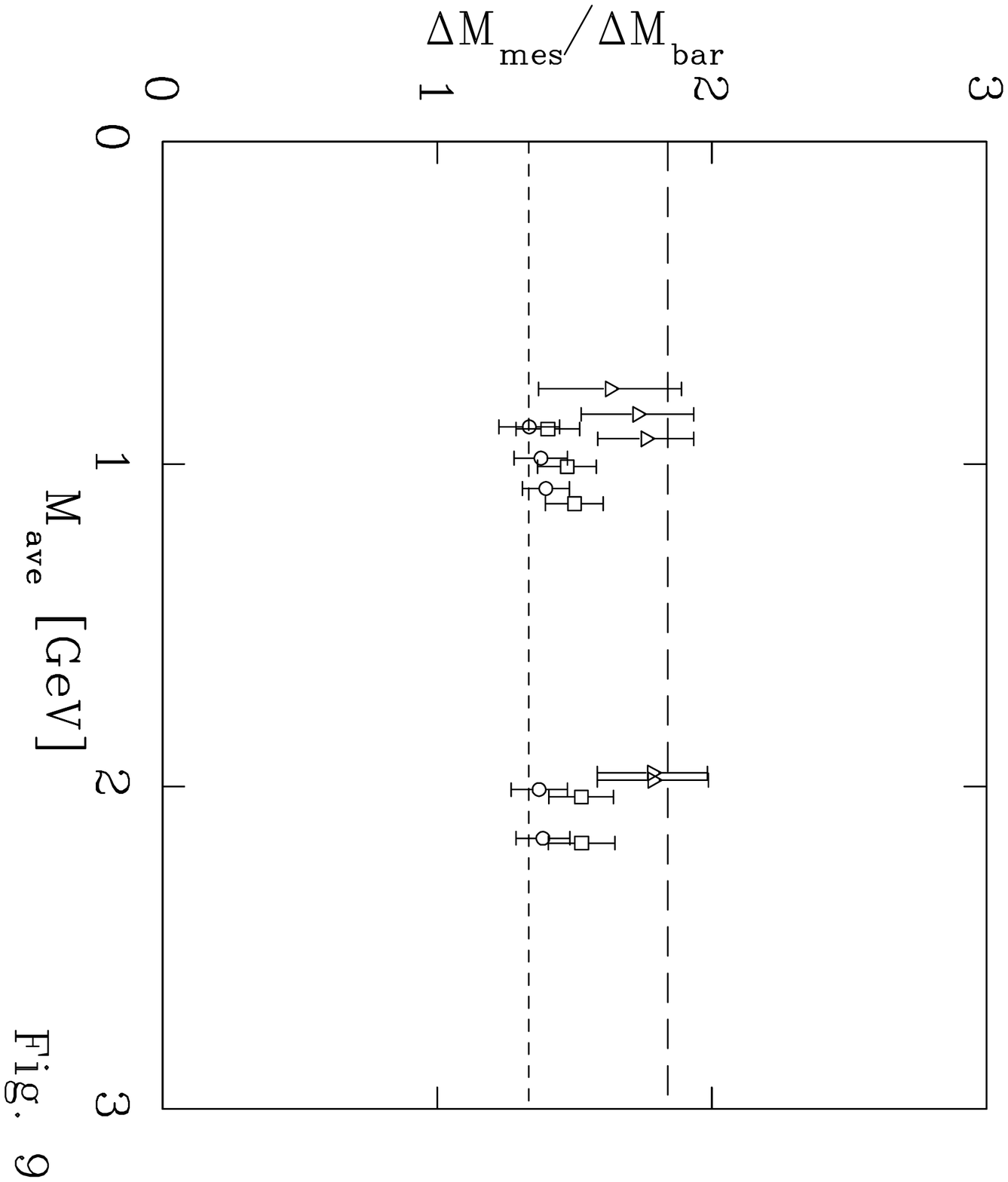}


\begin{thebibliography}{99}
\bibitem{bernard}For a review see C. Bernard, 
 Nucl. Phys. B(Proc. Suppl.) \textbf{94}, 159(2001).
\bibitem{koniuk}C. Stewart and R. Koniuk, 
 Phys. Rev. D \textbf{63}, 054503(2001).
\bibitem{manke}CP-PACS Collaboration, T. Manke \emph{et al.}, 
 Phys. Rev. D \textbf{62}, 114508(2000).
\bibitem{collins}S. Collins \emph{et al.}, 
 Phys. Rev. D \textbf{60}, 074504(1999).
\bibitem{UKQCD1}UKQCD Collaboration, K.C. Bowler \emph{et al}., 
 Phys. Rev. D \textbf{54}, 3619(1996).
\bibitem{cbary}R.M. Woloshyn, 
 Phys. Lett. B \textbf{476}, 309(2000). 
\bibitem{ali1}A. Ali Khan \emph{et al.}, 
 Phys. Rev. D \textbf{62}, 054505(2001).
\bibitem{alford}M. Alford, T.R. Klassen, and G.P. Lepage, 
 Nucl.Phys. \textbf{B496}, 377(1997).
\bibitem{spwave}R. Lewis and R.M. Woloshyn, 
 Phys. Rev. D \textbf{62}, 114507(2000).
\bibitem{leinweb1}D.B. Leinweber, R.M. Woloshyn, and T. Draper, 
 Phys. Rev. D \textbf{43}, 1659(1991).
\bibitem{valencia}R.M. Woloshyn, 
 Nucl. Phys. B(Proc. Suppl). \textbf{93}, 38(2001).
\bibitem{benmer}M. Benmerrouche, R.M. Davidson, and N.C. Mukhopadhyay, 
 Phys. Rev. C \textbf{39}, 2339(1989).
\bibitem{smear}C. Alexandrou, S. G\"usken, F. Jegerlehner, K, Schilling, 
and R. Sommer, 
 Nucl. Phys. \textbf{B414}, 815(1994).
\bibitem{pdg}D.E. Groom \emph{et al}., Eur. Phys. J. C \textbf{15}, 1(2000).
\bibitem{jess}CLEO Collaboration, C.P. Jessop \emph{et al}., 
 Phys. Rev. Lett. \textbf{82}, 492(1999).
\bibitem{kiselev}V.V Kiselev and A.K. Likhoded, hep-ph/0103169.
\bibitem{stringy}M. Ademollo, G. Veneziano, and S. Weinberg,  
 Phys. Rev. Lett. \textbf{22}, 83(1969).
\bibitem{lewellen}D.C. Lewellen, Nucl. Phys. \textbf{B392}, 137(1992).
\bibitem{beane}S.R. Beane, Phys. Rev. D \textbf{59}, 036001(1999).
\bibitem{frank}M. Frank and P.J. O'Donnell, 
 Phys. Lett. B \textbf{159}, 174(1985).
\bibitem{wise}See. for example, M.B. Wise, hep-ph/9805468.
\bibitem{mescalc}J. Hein \emph{et al.,} Phys. Rev. D \textbf{62}, 074503(2000);
 K.I. Ishikawa \emph{et al.}, Phys. Rev. D \textbf{61}, 074501(2000);
 P. Boyle, Nucl. Phys. B(Proc. Suppl.) \textbf{63A-C}, 314(1998).
\bibitem{mbcalc}CP-PACS Collaboration, S. Aoki \emph{et al.,} 
 Phys. Rev. Lett. \textbf{84}, 238(2000);
 UKQCD Collaboration, K.C. Bowler \emph{et al.}, 
 Phys. Rev. D \textbf{63}, 054506(2000);
 F. Butler \emph{et al.}, Nucl. Phys. B \textbf{430,} 179(1994).
\bibitem{barcalc}T. Bhattacharya, R. Gupta, G. Kilcup, and S. Sharpe, 
 Phys. Rev. D \textbf{53}, 6486(1996).
\bibitem{lipkin86}H.J. Lipkin, Phys. Lett. B \textbf{171}, 293(1986).
\bibitem{lipkin97}H.J. Lipkin and P.J. O'Donnell, 
 Phys. Lett. B \textbf{409}, 412(1997).
\bibitem{halzen}F. Halzen and A.D. Martin, \emph{Quarks \& Leptons} 
 (Wiley, New York, 1984)Chap. 2.
\bibitem{dgh}J.F. Donoghue, E. Golowich, and B.R. Holstein, 
\emph{Dynamics of the Standard Model} (Cambridge University Press, Cambridge, 
1992)Chap. 11.
\bibitem{savage90}M.J. Savage and M.B. Wise, 
 Phys. Lett. B \textbf{248}, 177(1990).
\end{thebibliography}
\end{document}